\documentstyle[12pt]{article}
\def\ll {\label}
\def\re{\ref}
\def\c{\cite}

\def\r1{(\ref{$1})}
\def\ot{\otimes}

\def\ti{\tilde}

\def\th{\theta}
\def\ba{\begin{array}{c}}

\def\sk{\smallskip}
\def\ea{\end{array}}

\def\ni{\noindent}
\def\si{\sigma}

\def\bet{\beta}
\def\ov{\over}

\def\l{\left}
\def\l({\left(}
\def\r){\right)}
\def\r{\right}
\def\rw{\rightarrow}

\def\la{\lambda}
\def\al{\alpha}

\def\be{\begin{equation}}
\def\bc{\begin{center}}
\def\ec{\end{center}}
\def\bit{\begin{itemize}}
\def\eit{\end{itemize}}
\def\ee{\end{equation}}
\def\ed{\end{document}}
\def\bea{\begin{eqnarray}}
\def\eea{\end{eqnarray}}
\begin{document}
\title{ Exactly integrable family  of generalized Hubbard models
with twisted Yangian symmetry}
\author { Anjan Kundu\footnote {e-mail: anjan@tnp.saha.ernet.in} \\
 Theory Division, Saha Institute of Nuclear Physics, 1/AF Bidhan Nagar,
\\700 064 Calcutta, India.} 
\maketitle
 \begin{abstract}
 {A strongly correlated electron system with controlled hopping, in the line
of the recently proposed generalized Hubbard models as candidates for high
$T_c$-superconductors, is considered. The model along with a whole class of
such systems are shown to be completely integrable with explicit quantum
$R$-matrices and the Lax operators.
 Inspite of  novelties in  the Bethe ansatz solution, the final results do
not deviate much from those of the standard Hubbard model.
 However, the symmetry of the
model is changed to a recently discovered twisted Yangian symmetry. }
\end{abstract}
 \ni {\it PACS numbers}: 04.20.Jb, 03.65.Fd, 71.10.Pm, 75.10.Jm

\sk

\ni {\it Key words}: Quantum integrability, Bethe ansatz, Generalized
Hubbard model with correlated hopping, twisted Yangian symmetry

\sk

\ni {\it Subject Classification}: Nonlinear Systems
\newpage
 
Some unusual but universal behavior of interacting fermion systems, especially
 in two dimensions, named as the Luttinger liquid theory  was
 speculated to be the basis for the high $T_c$-superconductivity
\c{hald81,ander88}.
 Most significant among such properties is the separation of charge and
 spin degrees of freedom.  Thus the charges of electrons are given
 to the pseudoparticle modes  { holons} and anti-holons
   with charges but without spins, while the spins are given to the
   {spinons} with spins but without any charge.  These are many-body
   collective modes with strong nonperturbative nature and unlike the
   quasiparticles of usual Fermi liquid which in the limit of vanishing
   interactions map into  free electrons, they can not exist without
   many-body interactions.  Such characteristic behavior of Luttinger liquid
   however is more common in one dimension and can be observed explicitly in
   Bethe ansatz solvable correlated electron models like Hubbard model
   \c{ovchin,holspin}. Due to this fact  the
   investigation of correlated electron systems in one-dimension has become
immensely important \c{ovchin,holspin,korepin,bariev}.

   Though the study of the standard Hubbard model itself  was identified to
   be promising \c{ander2}, more general models including higher nonlinear
   interactions with correlated hopping
are being proposed  \c{hirsch,SV,aa94} for better  description
   of the cuprate superconductors. Though ignoring certain terms and
restricting coupling parameters, some exact results were obtained in \c{SV}
and \c{aa94}, such generalized Hubbard models are not Bethe ansatz solvable
and clearly do not show complete integrability in one dimension.

Our aim is to propose a related though  different strongly correlated
electron model given by the Hamiltonian
 \bea H_\eta=
-\sum_{j,\si}c^\dag_{j (-\si)}c_{j+1(-\si)} 
 \{ t_{AA}+ (t^{ \si}_{AB}-t_{AA})(
 n_{j  (\si )}+n_{j+1 (\si )}) &+&(t_{AA}+ t^{ \si}_{BB}-2
 t^{ \si}_{AB})
  n_{j ( \si )}n_{j+1 ( \si )} \}\nonumber \\
&+& U \sum_j n_{i  (+)}n_{i (-)} +{\rm h.c}.
, \ll{Haa0}\eea
 where
$c^\dag_{j (\si)} (c_{j(\si)}) $
is the creation (annihilation) operator for an electron with spin $\si=\pm$
at site $j$ and  $ n_{j  (\si)}=c^\dag_{j (\si)} c_{j(\si)}  $ is the number
operator at site $j$ with $N_a$ being the total number of sites.
 The coupling constants are taken as 
$ t^\pm_{BB}= (t^{\pm}_{AB})^2= e^{\pm i2 \eta}$ 
and scaled  by putting $t_{AA}=1,$  where the  
  real parameters  $\eta$ and $U$ are kept arbitrary.
Notice that apart from the standard  Hubbard interaction represented by the
double occupancy term $H_U=U \sum_j n_{j  (+)}n_{j (-)} $ and
 the free hopping, the model (\re{Haa0}) contains additional interacting
terms influencing hopping. These are the
 Hirsch-like interaction \c{hirsch} given by the second term and higher
nonlinear interaction involving different sites given by the third term. In
particular, the hopping of up (down) spin electron is
 controlled by the presence of  down (up) spin electron at the same or at
neighboring sites. Moreover, since
 the coupling constants  are different for up and down spins as well as for
left and right hoppings, their hopping rates might be different.  It may be
mentioned that the model proposed in \c{aa94} is of the form
 (\re{Haa0}) with coupling constants independent of spins: $t^{ \si}_{AB}
=t_{AB}, t^{ \si}_{BB}=t_{BB}.$

We show that, as opposed to the model of \c{aa94} the generalized
 Hubbard model  (\re{Haa0}) is exactly solvable by the Bethe ansatz with
  unusual features. Moreover, the corresponding Bethe ansatz results do not
deviate much from those of the standard
 Hubbard model and  proceeding analogously one can expect to find
similar Luttinger liquid like behavior.
   However, the present model exhibits a new type of symmetry
  discovered very recently, namely the
twisted Yangian symmetry at the infinite chain limit. Furthermore, 
along with a whole class  it
  belong to the  completely integrable quantum systems with sufficient
 number of independent conservation laws. The associated $R$-matrix and the
 Lax operators satisfying the quantum Yang-Baxter equation can be extracted
 in the explicit form, which are found to be intimately related with those
 of the Hubbard model.

 Using the fermionic property
$ (n_{j (\pm)})^2= n_{j (\pm)}, $ it is convenient to rewrite Hamiltonian
(\re{Haa0}) with our choice of parameters in the form 
\be
 H_\eta= -\sum_{j} c^\dag_{j+1 (+)}c_{j (+)}
e^{i \eta [ n_{j (-)}+ n_{j+1 (-)}]}
+c^\dag_{j+1 (-)}c_{j (-)}
e^{-i \eta [ n_{j (+)}+ n_{j+1 (+)}]}  
+ U  n_{j  (+)}n_{j (-)}  +{\rm h.c}.
 \ll{Haa1}\ee
 We show below that the eigenvalue problem of the model
 can be solved exactly by using the coordinate formulation of the Bethe
 ansatz. Though this method proposed first by Bethe \c{bethe} has been
applied since to a number of models and by now has become almost
 an algorithmized  problem, the present model  shows surprises
  in its Bethe ansatz solution.

This subtle feature is manifested already in the simple two-particle case,
  if we consider the wave function involving one up and one down-spin
electrons created
 respectively at the sites $x_1$ and $x_2$.  We have to distinguish
naturally between the state $\psi^{(+-)}(x_1,x_2)$ in the sector $x_1<x_2 $,
i.e when the up-spin is created left to the down-spin and the state
$\psi^{(-+)}(x_1,x_2)$ in $x_1>x_2 $
 with the up-spin to the  right of the down-spin. Since the interactions are
short-ranged appearing only for the opposite spins occupying the same or the
nearest-neighbor sites, all interactions naturally vanish when spins are
placed well apart.

Therefore for $x_1, x_2$ far apart the wave functions $\psi^{(\pm
\mp)}(x_1,x_2) \ $ should satisfy the free discrete Schr\"odinger equation
\bea
 \psi^{(\pm \mp)}( x_1-1,x_2)+\psi^{(\pm \mp)}( x_1+1,x_2)
  &+& \psi^{(\pm \mp)}( x_1,x_2-1)   
+ \psi^{(\pm \mp)}( x_1,x_2+1) 
\nonumber \\ &=&  -E \psi^{(\pm \mp)}( x_1,x_2).
\ll{bethe0} \eea
On the other hand for the nearest neighbor occupation of the opposite spins 
   the interactions of the hopping terms come into play, resulting for
  $x_2=x_1 + 1, x_1=x ,$ the Schr\"odinger equation
 \bea
 \psi^{(+-)}( x-1,x+1)+\psi^{(+-)}( x+1,x+1)e^{-i \eta}
  &+& \psi^{(+-)}( x,x)e^{-i \eta} +\psi^{(+-)}( x,x+2) \nonumber \\ &=& - E
\psi^{(+-)}( x,x+1). \ll{bethe01} \eea 
and similarly for $x_2=x_1 - 1, x_1=x
,$ related to $ \psi^{(-+)}.$ Now as the Bethe ansatz demands,
 the same solution $ \psi^{(+-)}$ must hold for both these equations, i.e.
eqn. (\re{bethe0}) at the limit $x_2 \rw x_1+1$ should formally
 coincide  with  (\re{bethe01}). This leads to the unexpected result \be
\psi^{(+-)}(x_1,x_2)\mid_{x_1 \rw x_2 }= e^{- i \eta} \psi(x_2,x_2), \ \
\mbox{ and similarly} \ \ \psi^{(-+)}(x_1,x_2)\mid_{x_1 \rw x_2 }= e^{i
\eta} \psi(x_2,x_2), \ll{limit} \ee
 where $ \psi(x,x) \equiv \psi^{(+-)}(x,x)= \psi^{(-+)}(x, x) $
 at the coinciding points.  This unusual {\em anyonic} type feature
introduced through coupling constant $\eta$ can be understood also at the
operator level, as will be demonstrated below.

We notice further that at $x_2=x_1=x$ both interactions involving hopping
 terms with parameter $\eta$ as well as the Coulomb term with coefficient
 $U$     become active resulting the 
Schr\"odinger equation
\bea
 \psi^{(+-)}( x-1,x)e^{i \eta}+\psi^{(-+)}( x+1,x)e^{-i \eta}
  &+&
\psi^{(-+)}( x,x-1)e^{-i \eta}
+\psi^{(+-)}( x,x+1)e^{i \eta}
\nonumber \\+ U \psi( x,x) &=& - E \psi( x,x).
\ll{bethe02} \eea
Demanding again that   (\re{bethe0}) should be compatible with
 (\re{bethe02}) when $x_1 \rw x_2$ and using the relations (\re{limit}) at
 the coinciding points we arrive at the consistency condition
 \bea
 \psi^{(+-)}( x+1,x)-   
 \psi^{(-+)}(x+1,x) e^{-2i \eta}  &+& 
\psi^{(+-)}(x,x-1) -
\psi^{(-+)}(x,x-1) e^{-2i \eta}
\nonumber \\ &-& U  \psi^{(+-)}(x_1 ,x_2)\mid_{x_1 \rw x_2=x} = 0,
\ll{bethe03} \eea
Defining now the wave functions in  the standard   Bethe ansatz form
\be
 \psi^{(\pm \mp)}( x_1,x_2)= A^{\pm \mp}_{p_1p_2} e^{i(p_1x_1+p_2 x_2)}
-
  A^{\pm \mp}_{p_2p_1} e^{i(p_2x_1+p_1x_2)},
\ll{bawf}\ee
which is valid basically in the sectors
$x_1 \not = x_2,$
one can easily calculate the energy eigenvalue from (\re{bethe0}) as
$E=- 2(\cos p_1+ \cos p_2).$ 
To evaluate  (\ref{bawf}) at the coinciding points,
 one has to consider (\re{limit})  yielding
 $\psi^{(+-)}(x_1,x_2)\mid_{x_1 \rw x_2}
= e^{-2 i \eta} 
\psi^{(-+)}(x_1,x_2)\mid_{x_1 \rw x_2},$ which in turn  
 leads to the nonstandard  relation
\be 
A^{+-}_{p_1p_2} 
-
  A^{+-}_{p_2p_1}
= e^{-2 i \eta} (A^{-+}_{p_1p_2} 
-
  A^{-+}_{p_2p_1}). \ll{aa} \ee 
Inserting  ansatz (\ref{bawf}) in the
consistency condition (\re{bethe03}) and using  relation (\ref{aa}) one
gets finally the two-particle scattering matrix
 \be S(\la_1^0-\la_2^0)=
{(\la_1^0-\la_2^0) \Sigma+ i {U \over 2} \hat P \over \la_1^0-\la_2^0 +i {U
\over 2}} \ll{smatrix}\ee
 defined as $ \ \ A^{ab}_{p_2p_1}= \sum_{cd}
S(\la_1^0-\la_2^0)^{ba}_{cd} \ \ \ A^{cd}_{p_1p_2} \ ,$ where the obvious
relations like $ A^{\pm \pm}_{p_2p_1}=
 A^{\pm \pm}_{p_1p_2} $ are imposed. 
Here $ \lambda_i^0=\sin p_i,$
 $ \hat P ={1 \over 2} \left( I+ {\vec \sigma }\otimes { \vec \sigma}
\right) $ is the permutation operator and 
\be
 \Sigma= diag (1, 
 e^{2i \eta},  e^{-2i \eta}, 1).
\ll{Sigma} \ee

It is crucial to note  that the $S$-matrix (\ref{smatrix}) satisfies
the well known Yang-Baxter equation \c{dvega}
\be
S_{12}(\la_1-\la_2)S_{13}(\la_1-\la_3)S_{23}(\la_2-\la_3)=S_{23}(\la_2-\la_3)
S_{13}(\la_1-\la_3)S_{12}(\la_1-\la_2), 
\ll{ybe} \ee
representing
 the factorizability condition for the many-particle scattering into the
two-particle ones. This enables us to solve the general $N$-body
  problem  through the Bethe ansatz,
 when the state  corresponds   to the presence of 
 total $N$  number of  electrons with $M$ down-spin electrons.  Each particle
scattering through the rest and returning to its original position would
generate a string of $S$-matrices (\re{smatrix}) in the factorized form. On
the other hand,
 for a closed chain with the periodic boundary condition this would be
equivalent to the
 shift operator   over the total  number of lattice sites  $N_a$. 
Diagonalizing this relation, as seen from (\re{smatrix}) one gets
 \be
e^{ip_j N_a}= e^{-i2\eta M} \prod_{\al=1}^M { \la_\al -\la_j^0 +i{U \over 2}
\over \la_\al -\la_j^0 }, \ll{betheeqn1} \ee
   along with the relations
 \be
 e^{-i2\eta N} \prod_{j=1}^N { \la_\al
-\la_j^0 +i{U \over 2} \over \la_\al -\la_j^0 } =\prod_{\bet=1}^M { \la_\al
-\la_\bet +i{U \over 2} \over \la_\al -\la_\bet -i{U \over 2}}.
\ll{betheeqn2} \ee
 Taking logarithm of these  Bethe equations we get
 \bea p_j N_a
&=& 2 \pi I_j +2 \sum_{\al=1}^M \tan ^{-1} ({4 \ov U} ( \la_\al -\sin p_j))
{-2\eta M}\nonumber \\ 2 \sum_{j=1}^N \tan ^{-1} ({4 \ov U} ( \la_\al -\sin
p_j)) &=& 2 \pi J_\al + 2 \sum_{\bet=1}^M \tan ^{-1} ({2 \ov U}
 ( \la_\al -\la_\bet)) {+2\eta N}
\ll{betheeqn3} \eea
where $I_j, J_\al$ are integers or half odd integers.

We define { charge} and { spin} rapidities as  $p_j (q_j),
j=1,2,\ldots ,N$ and
$\la_\al (\rho_\al), \al=1,2,\ldots, M $,
respectively,  where $q_j={2 \pi \over N_a } I_j$  and $
 \rho_\al={2 \pi \over N_a} J_\al$ are given through two  independent
sets of quantum numbers $ I_j $ and $ J_\al.$ 
 Comparing (\re{betheeqn3}) with the Bethe ansatz results of the Hubbard
model \c{wuliebh} we may conclude that, though due to the inclusion of
various interacting terms influencing hopping  our generalization
  differs considerably from the standard Hubbard model,
 the final Bethe ansatz results do not show significant changes. 
Nevertheless, though the energy eigenvalue is obtained in the same form, the
  determining equations for $p(q)$ and $ \la(\rho)$ are modified due to the
appearance of additional phases involving coupling constant $\eta$ and
particle numbers $N$ and $M$. Going  to the thermodynamic limit,
     analogous to the Hubbard model \c{ovchin,holspin,hubexi}, one expects
     to show that all low lying excitation modes of the system
 are expressed through the decoupled charge and spin degrees of freedom.

 Interestingly, for our choice of the coupling constants, the model
 (\re{Haa0}) not only becomes Bethe ansatz solvable, as we have seen above,
but also turns out to be a completely integrable quantum system with higher
conservation laws like the original Hubbard model \c{wadh}.
 The $R^{\rm Hub}(\la, \mu)$-matrix and the Lax operator $L^{\rm Hub}(\la)$
of the Hubbard model as a solution of the quantum Yang-Baxter equation
\c{dvega} 
\be
R_{12}(\la_1,\la_2)L_{1j}(\la_1)L_{2j}(\la_2)=L_{2j}(\la_2)
L_{1j}(\la_1)R_{12}(\la_1,\la_2)
\ll{qybe} \ee
 were given in a convenient form in \c{wad95} as
\be
L^{\rm Hub}_{aj}(\la_a)= (L^{\si_{(+)}}_{aj}(\la_a)\ot 
L^{\si_{(-)}}_{aj}(\la_a)) \exp (h_a \si^3_{(+)a}\si^3_{(-)a})
\ll{lhub}\ee
and
\bea
R^{\rm Hub}_{12}(\la_1,\la_2)&=&  [ \cos \ti \th_{12} \cosh h_{12}
(L^{\si_{(+)}}_{12}(\th_{12})\ot 
L^{\si_{(-)}}_{12}(\th_{12}))\nonumber \\ &+&
\cos  \th_{12} \sinh h_{12}
(L^{\si_{(+)}}_{12}( \ti \th_{12})\ot 
L^{\si_{(-)}}_{12}(\ti \th_{12}))
  (\si^3_{(+)1}\si^3_{(-)2})]
\ll{rhub}\eea
where $L^{\si_{(\pm)}} $ correspond to the $6$-vertex  free-fermionic
model, $h_{12}=h_1- h_2 $ and $\th_{12}=\la_1-\la_2, \ \
 \ti \th_{12}=\la_1+\la_2$
represent the dependence on the difference and the sum of the spectral
parameters. The Pauli matrices ${\si_{(\pm)}} $ correspond to the 
 spin up/down fermion operators $c_{(\pm)}$  and the parameters $h_a$ are
defined as $ \sinh 2h_a = {U\ov 4} \sin 2 \la_a$ \c{wad95}. The Lax-operator
and the $R$-matrix
  for  the present model  are linked intimately   with those of the Hubbard
model and can be obtained easily from (\re{lhub}) and (\re{rhub}) through
 { twisting} transformation:  
\be
R_{12}(\eta, \la_1, \la_2)=F_{12} (\eta) R^{\rm Hub}_{12}(\la_1,\la_2)
 F_{12}(\eta); \ \   
L_{aj}(\eta,\la_1)=F_{aj}(\eta) L^{\rm Hub}_{aj}(\la) F_{aj}(\eta)
,\ll{RL}\ee
 where  the twist operator is given by \be F_{aj}(\eta)= e^{i \eta
(\si^3_{(-)a} \si^3_{(+)j} - \si^3_{(+)a} \si^3_{(-)j} )}. \ll{F} \ee
 Recall
that the twisting transformation \c{reshet} generates new $R$-matrix and
$L$-operator solutions exploiting a nontrivial symmetry of the Yang-Baxter
equation.

Another important issue is to study the effect of the additional terms is
changing the symmetry of this model. It is well known that the Hubbard model
exhibits a Yangian symmetry in the infinite chain limit \c{kory,gohomany}.
It is therefore intriguing to ask whether the present generalization
destroys the original symmetry completely or deforms it to another one.
 The origin of the Yangian symmetry in the Hubbard model is  the rational
$R$-matrix of the $XXX$ spin chain embedded in it,
 which gets associated with the algebra of the monodromy matrix at the
infinite interval \c{gohomany}. The corresponding rational $R$-matrix
 here is its  twisted version (\re{smatrix}), which gives
 the present model  a new type of twisted Yangian symmetry, discovered
recently \c{biruty,kulishty}. The expansion of the monodromy matrix: $ \ \
T^{\alpha \beta}= \tau^{\alpha \alpha} \delta _{\alpha \beta} +
\sum_{n=o}^\infty {\tilde t^{\alpha \beta}_{(n)} \over \lambda ^{n+1}} \ ,
\alpha ,\beta =1,2 \ \ $ yields
 generators $\tau^{\alpha \alpha}, \tilde t^{\alpha \beta}_n $
of the infinite dimensional twisted Yangian 
algebra $Y_\eta (gl_2)$, defining 
 relations of which are given in explicit form in \c {biruty}. We can find a
realization of this algebra through the fermionic operators by expressing
first the generators as 
\bea \tilde t^{\alpha \beta}_{(n)}&=&\sum_{j}
\tau^{\alpha \alpha}_{j-}
 (\tau^{\alpha \alpha}_{j})^{{1 \over 2}} 
t^{\alpha \beta}_{j(n)}(\tau^{\beta\beta}_{j})^{{1 \over 2}}
\tau^{\beta\beta}_{j+} 
, \ \ \  \mbox {for}\ \  \alpha \not = \beta, \nonumber \\
  \tilde t^{\alpha \alpha}_{(n)}&=& \tau^{\alpha \alpha}t^{\alpha \alpha}_n{(n)} 
\ll{tyangopr}\eea
with the  notations 
$ \
\tau^{\alpha\alpha}_{j \pm}\equiv \prod_{k>j (k<j)} 
\tau^{\alpha\alpha}_k
,$ and $ \ \tau^{\alpha \alpha}= \prod_{k}\tau^{\alpha \alpha}_{k}.$ 
Note that for $\tau^{\alpha \alpha}_k=1$ the expressions (\ref{tyangopr})
reduce to the undeformed $Y(sl_2)$ Yangian generators
 $t^{\alpha \beta}_{(n)}= \sum_j t^{\alpha \beta}_{j(n)}.$ Therefore using
 the
  well known Yangian representation for the Hubbard model \c{kory} involving
$c^\dagger_{j (\pm)}, c_{j (\pm)} $ and $ n_{j(\pm)}= c^\dagger_{j
(\pm)}c_{j (\pm)}, $ along with
  additional expressions 
\be
\tau^{11}_j= e^{2i \eta n_{j(-)}}, \ 
\tau^{22}_j= e^{-2i \eta n_{j(+)}}, \ \  \mbox{and} \ \   
t^{11}_{j(0)}= n_{j(-)},\  t^{22}_{j(0)}= n_{j(+)},   \ll{tau}\ee 
we can obtain an exact representation of the twisted Yangian in
 fermion operators. In the line of \c{kory}, it can be shown now by direct
check that the Hamiltonian (\ref{Haa1}) for infinite chain commutes with the
generators of this twisted algebra along with a complementary set of such
generators  obtained by replacing 
\be c_{j(-)} \rw
c_{j(-)} e^{-2i \eta j}, \ \ \ \ c_{j(+)} \rw (-1)^j c^\dagger_{j(+)}, \ \
\mbox { with } \ U \rw -U, \ \eta \rw -\eta, \ll{dual} \ee
  since under (\ref{dual}) the Hamiltonian remains invariant.
  This   proves   a novel twisted $\ Y_\eta (gl_2) \oplus Y_{-\eta}(gl_2) \
$ Yangian symmetry for our generalized Hubbard model $H_\eta$ (\ref{Haa1})
at the infinite chain limit. It is easy to check that at $\eta=0$ one
recovers the result related to the standard Hubbard model \c{kory}. 

Finally we mention about a possibility of extending the present model
 to an one-parameter family of integrable models by introducing
an additional coupling constant $g$ in the form
\bea
 H_{\eta g}=- \sum_{j} ( c^\dag_{j+1 (+)}c_{j (+)}
e^{i[(\eta -g) n_{j (-)}+(\eta +g) n_{j+1 (-)}]}
&+& c^\dag_{j+1 (-)}c_{j (-)}
e^{-i[(\eta +g) n_{j (+)}+(\eta -g) n_{j+1 (+)}]})  \nonumber \\
&+& U  n_{j  (+)}n_{j (-)}   + {\rm h.c.} ,
 \ll{Haa2}\eea
which reduces to  $H_\eta$ (\re{Haa1}) at $g=0$ and generates  
at $g= \pm \eta$   new type of
models like
\be
 H_{\eta +}= -\sum_{j}  c^\dag_{j+1 (+)}c_{j (+)}
e^{i2 \eta  n_{j+1 (-)}}
+c^\dag_{j+1 (-)}c_{j (-)}
e^{-i 2 \eta n_{j (+)}}
+ U  n_{j  (+)}n_{j (-)}   + {\rm h.c.} 
 \ll{Haa3}\ee
Note that these models are different from those proposed  
 in \c{bariev} or \cite{kusakabe}.
Remarkably,  though the coupling constants 
in (\ref{Haa2}) are  given as combinations of 
$\eta \pm g$, the eigenvalues as well as 
 the scattering matrix remain independent of the parameter $g$. This family
of models producing the same Bethe ansatz results and sharing the same
symmetry can be represented by
 the $R$-matrices and the Lax operators, which may be obtained from
  those with   $g=0$ through a simple gauge transformation
\be
R_{12}(\eta, g, \la_1, \la_2)=A_{12} (g) R_{12}(\eta,  \la_1, \la_2)
 A^{-1}_{12}(g); \ \   
L_{aj}(\eta, g,\la_1)=A_{aj}(g) L_{aj}(\eta,\la) A_{aj}^{-1}(g)
\ll{RLA}\ee
  as an operator dependent similarity transformation through
  $ 
A_{aj}(g)= e^{i g (\si^3_{(-)a} \si^3_{(+)a} + \si^3_{(+)j} 
\si^3_{(-)j} )}. $

The present  study of the generalized Hubbard model shows that additional
terms influencing hopping with resemblance to the recently proposed
 models for high $T_c$ superconductors, may be introduced retaining its
integrability in one dimension with nearest-neighbor interactions.
Moreover, for a change in operators as
 \be c^\dag_{j (\pm)} \rw \tilde
c^\dag_{j (\pm)}= e^{\pm i \eta n_{ j (\mp)}} c^\dag_{j (\pm)} ,\ \ \ c_{j
(\pm)} \rw \tilde c_{j (\pm)}= e^{\pm i \eta n_{ j (\mp) }} c_{j (\pm)},
\ll{anyop} \ee 
the generalized model (\re{Haa1}) can be transformed back
 to the original form of the Hubbard model. However the resultant operators
would show not free fermionic but { anyonic} type of commutation relations
 \be \ti
c^\dag_{j (+)} \tilde c^\dag_{j (-)}+ e^{2i \eta} \ti c^\dag_{j (-)} \ti
c^\dag_{j (+)}=0 , \ \ \{ \ti c^\dag_{j (\pm)}, \tilde c_{j (\pm)} \}= e^{\pm 2 i
\eta n_{ j (\mp)}} \ll{anyal} \ee
 etc.
We have witnessed the reflection of this intriguing  feature  
  in the Bethe ansatz procedure with the wave functions suffering 
  phase jumps at the boundaries of two different sectors.  Usually in the Bethe
 ansatz, the matching of the wave functions at the boundaries is assumed.
However, the present model suggests for a more careful comparison with the
consideration of their phase factors.

 Inspite of many nontrivialities involved in the generalized model, the
final Bethe ansatz results are not much different from the original Hubbard
model, apart from a symmetry change and  modification in the Bethe
equations. This effect is like putting the corresponding vertex model in
vertical and horizontal electric fields, which spoils the invariance of the
Boltzmann weights under inversion of all arrows, as happens in the
asymmetric
 $6$-vertex model  \c{yansud} or  the Dzyaloshinsky-Moriya interaction
\c{dm}. Only in the present model the fields are not external but caused by
the interaction of different kinds of bonds.
 The original Yangian symmetry of the Hubbard model is deformed to the
twisted Yangian symmetry, while the quantum $R$-matrix, Lax operators etc.
are related through twisting transformation.

 Nevertheless, like the twisted Heisenberg spin chain \c{suther}
 or the  Hubbard model  with Ahronov-Bhom period \c{kusakabe} such
interactions, as shown in a recent work \c{shastry97}, can also be absorbed
in the boundary conditions, though in a more involved way. This also makes
the investigation of some important problems, like
 the influence on the effective period \c{kusakabe},  changes in the
 finite temperature behavior \c{klubar} and  related conformal
properties   \c{wor,klubat91,kar}  and also the modification of
  correlation functions worth studying.  Similar idea can also be used
for generating integrable coupled anisotropic spin chains \c{kunpr}.
Following a different approach an integrable coupled spin chain and
 a quasi two dimensional  extension of the Hubbard model had  been
obtained in some earlier works \c{borov93,zvyag}.

The author thanks Dr. Indrani Bose for valuable discussions and the
 referee for his constructive comments.


\newpage

\begin{thebibliography}{99}
\bibitem{hald81} Haldane , J. Phys. C 14 (1981) 2535
\bibitem{ander88} P. W. Anderson and Z. Zhau, Phys. Rev. B 37 (1988) 627

P. W. Anderson, Phys. Rev. Lett. 64 (1990)  1839

P. W. Anderson, {\it The theory of superconductivity in the high $T_c$
cuprates}, (Priceton Univ. Press, Princeton, 1997)
\bibitem{ovchin} J. Carmelo and A. Ovchinnikov, J. Phys. C3 (1991) 757

J. Carmelo, P. Horsch,P. A. Bares and  A. Ovchinnikov, Int. J. Mod. Phys. B
5 1991) 3
\bibitem{holspin} H. Frahm and V. E. Korepin, Phys. Rev. B 42 (1990) 10553;
{\it ibid}, Phys. Rev. B 43 (1991) 5653

F. H. L. Essler and V. E. Korepin, Phys. Rev. Lett. 72 (1994) 908;
  {\it ibid},   Nucl.
Phys. B 426 (1994) 505
 \bibitem{korepin} G. Albertini, V. E. Korepin and A.
Schadschneider, J. Phys. A 28 (1995) L303 
\bibitem{bariev} R. Z. Bariev, J.
Phys. A24 (1991) L549

R. Z. Bariev, A. Kl\"umper, A. Schadscheider and J. Zittartz, Phys. Rev. B
50 (1994) 9676
\bibitem{ander2} P. W. Anderson,  Phys. Rev. Lett. 64 (1990) 1839
\bibitem{hirsch} J. Hirsch, Phys. Lett. A 134 (1989) 451; 

J. Hirsch, Physica 158C (1989) 326
\bibitem{SV} R. Strack and D. Vollhardt, Phys. Rev. Lett. 70 (1993) 2637

A. A. Ovchinnikov, Mod. Phys. Lett. 7 (1993) 1397
\bibitem{aa94} L. Arrachea and Aligia,
 Phys. Rev. Lett. 73 (1994) 2240
\bibitem{bethe} H. Bethe, Z. Phys. 71  (1931) 205
\bibitem{wuliebh} E. H. Lieb and F. Y. Wu,  Phys. Rev. Lett. 20 (1968) 1445
\bibitem{hubexi} C. F. Coll, Phys. Rev. B 9 (1974) 2150
\bibitem{reshet} M. Wadati, T. Deguchi and Y. Akutsu, Phys. Report,  180
(1989) 247

N. Reshetikhin, Lett. Math. Phys. 20 (1990) 331

\bibitem{wadh}
  B. S. Shastry,  Phys. Rev. Lett. 56 (1986) 2453; J. Stat.
Phys. 50 (1988) 57

E. Olmedilla and M. Wadati, Phys. Rev. Lett. 60 (1988) 1595

\bibitem{dvega} 
H. J. de Vega , Int. J. Mod. Phys. A 4 (1989)
2371

\bibitem{wad95} M. Shiroishi and M, Wadati, J. Phys. Soc. Jpn. 64 (1995)
57 

\bibitem{kory} D. B. Uglov and  V. E. Korepin, Phys. Lett. A190 (1994) 238  
\bibitem{gohomany} S. Murakami and F. G\"ohmann,
 Phys. Lett. A 227 (1997) 216

 F.  G\"ohmann and V. Inozemtsev, Phys. Lett. A 214 (1996) 161

\bibitem{biruty} B. Basumallick and P. Ramadevi, Phys. Lett. A 211  (1996)
339
\bibitem{kulishty} A. Stolin and P. Kulish,  {\it New rational solution of
Yang-Baxter equation and deformed Yangians} q-alg/9608011
\bibitem{yansud} B. Sutherland, C. N. Yang and  C. P. Yang,
 Phys. Rev. Lett. 19 (1967) 588
\bibitem{dm}
F. C. Alcaraz and W. F. Wreszinski, J. Stat. Phys. 58 (1990) 45
\bibitem{klubar}
J. Suzuki, Y. Akutsu and M. Wadati, J. Phys. Soc. Jpn 59 (1990) 2667


 A. Kl\"umper and R. Z. Bariev, Nucl. Phys. B 458 [FS] 
 (1996) 623
\bibitem{suther}
S. Shastry and B. Sutherland, Phys. Rev. Lett., 65 (1990) 243;

B. Sutherland, Phys. Rev. Lett., 74 (1995) 816

\bibitem{shastry97} H. J. Schulz and B. S. Shastry, 
Phys. Rev. Lett, 80 (1998) 1924

\bibitem{kusakabe}
K. Kusakabe, J. Phys. Soc. Jpn., 66 (1997) 2075
\bibitem{wor} F. Woynarovich, J. Phys. A 22 (1989) 4243:

C. Destri and H. J. de Vega, Phys. Rev. Lett. 69 (1992) 2313
\bibitem{klubat91}
A Kl\"umper, M. T. Batchelor and P. A. Pearce, J. Phys. A 24 (1991) 3111
\bibitem{kar} M. Karowski, Nucl. Phys. B300 [FS22] (1988) 473
\bibitem{kunpr} Anjan Kundu, {\it Quantum integrable systems: construction,
solution, algebraic aspect}, hep-th/9612046
\bibitem{borov93} A. E. Borovik, S. I. kulinich, V. Yu. Popkov and Yu. M.
Strzhemechny, Phys. Lett. A 174 (1993) 407 
 \bibitem{zvyag} 
A. A. Zvyagin, Sov, J. Low Temp. Phys. 18 (1992) 723

\end{thebibliography}
 \end{document}